# Quantum kinetic theory: correlations and linking.

J. H. Hannay, H. H. Wills Physics Laboratory, University of Bristol, Tyndall Ave, Bristol BS8 1TL, UK.

*Abstract*
Classically the kinetic theory for a perfect gas has zero spatial number density correlation between separate points because the particles are independent. But the joint spatial and temporal correlation is non-zero (and easily calculable) because each individual particle moves in a straight line. The same holds for particle flux density correlation. The equivalent 'quantum kinetic theory' correlations are evaluated here via Feynman paths with their direct access to geometry and topology. The calculation is exact, yielding known special functions, but it is quite primitive physically. No heat bath, and no multi-particle statistics are invoked (the gas is thus 'Boltzmann'). Formally it reduces to path analysis of Brownian motion, in fact, of Brownian loops (suitably analytically continued). A check of the results is their correct classical limit. Attention is paid to the all-time-integral of the flux density correlation, with its topological significance. A particle's random path, in the presence of a fixed hypothetical hoop of arbitrary shape, accumulates a random linking number by repeatedly passing through it, positively or negatively. The mean is zero and the mean square is infinite, uninformatively. The interest lies in two separate arbitrary hoops: the two linking numbers are correlated, their average product grows linearly, and is calculated from the flux density correlation. For a charged particle, this would produce correlation in the induced Ampère magnetic circulation around the hoops.

*Introduction*

In the classical kinetic theory of a perfect gas at temperature *T*, the correlation of particle number densities at separate positions at the same time is of course zero, because the particles are independent. But if the times are separate too, the correlation is not zero because each individual particle moves (at constant velocity) from one place to another, and the correlation is easy to calculate (5). A feature of richer interest is the particle flux density. This is zero on average, of course, but as for the particle number density, there is non–zero correlation between the flux densities at separate positions and times. This correlation (by definition) supplies the average product of two particle fluxes: that through one arbitrarily chosen hypothetical infinitesimal area element in space, and that through any other, with any specified time delay. It is fairly straightforward to calculate this (20); as throughout this account, it is a single particle effect.

Pursuing this classical particle flux density correlation in a perfect gas, the flux densities can be generalized to fluxes. Instead of an infinitesimal area element one considers a hypothetical finite size piece of surface, arbitrarily curved, and fixed in space. The average rate at which particles cross this surface (counting backwards as negative) is zero, but for two such arbitrary surfaces there is correlation between the two fluxes at any chosen delay. Perhaps the most interesting feature of this correlation is not its full detail, but the integral of the flux correlation over all delay times, because this is topological. It depends only on the shape of the boundary 'hoops' of the two pieces of surface (as well as their locations), and can be found from the flux densities by integration using Stokes' theorem. Passing through the hoops in the opposite order contributes equally, so often, as in the reinterpretation in the next paragraph, a factor of two is needed.

Expressed differently; over a very long duration of a single particle's trajectory it will pass through or 'link with' each of the two hoops a number of times (positively or negatively). The product of these two linking numbers divided by the duration equals the time integrated flux correlation in the long duration limit. For $N$ particles this is multiplied by $N$. If the particles have charge $q$ (but are still assumed non-interacting) then each passage of the particle through a hoop generates a pulse of magnetic circulation $\oint \mathbf{B}(\mathbf{r},t) \cdot d\mathbf{r}$ by Ampère's law, the total time integral of this being $q\mu_0$. The calculation thus supplies the time-integrated correlation of the two circulations.

Here, this classical account is to be rendered quantum, 'quantum kinetic theory', the quest being to find the correlations and the consequent linking product for a perfect gas. It is natural to use Feynman's path integral prescription of Schrödinger quantum mechanics [Feynman and Hibbs 1965, Feynman 1972]. For ordinary quantum mechanics involving time but not temperature, the (complex) weights of the different paths are phase factors (imaginary exponents). For ordinary quantum statistical mechanics, for example for the partition function, that involve temperature but not time, the weights are real (real exponents). For the present problem of correlations both temperature and time are involved, but the analysis in such cases is not really harder, being accessible analytically from the purely real case now described.

For free motion, the relevant (exact) path analysis is formally equivalent to that of Brownian motion, that is, of diffusion. The formality perhaps needs emphasising: proper Brownian motion is of course classical, but here it is a formal trick for doing (thermal) quantum mechanics, the classical motion is in straight lines, quite different from Brownian motion. The path integral is that which solves the diffusion equation $D\nabla^2 Q = \partial Q / \partial t$ with a standard, but dramatic, reinterpretation of the meaning of all the variables. The diffusion coefficient $D$ becomes an absolute constant of the correct physical dimensions length$^2$/time. The 'time' variable $t$ is not true time at all, but a fake time – a mere parameter of the dimensions of time, proportional to the reciprocal of temperature $T$.

$$D = \hbar/2m, \qquad t = \hbar/kT \tag{1}$$

The diffusion equation describes how the quantity $Q$ abstractly 'evolves' under change of this parameter. Roughly speaking, the value of the parameter needs to be 'evolved' from zero (infinite temperature, where heat is entropy-free and mimics work, and mechanics applies) to the value appropriate to the desired finite temperature of the thermal problem in question. The quantity $Q(\mathbf{r})$, normally a probability density (for $\mathbf{r}$ at time $t$) in the diffusion equation, now represents a rather more complicated real scalar. Specifically, (though this will not be further needed), it is the (unnormalised) positional quantum thermal density matrix $\rho(\mathbf{r},\mathbf{r}',t) = \langle \mathbf{r}|\exp(-H/kT)|\mathbf{r}'\rangle$; for a free particle this depends only on $\mathbf{r} - \mathbf{r}'$, so one can take $\mathbf{r}'=\mathbf{0}$. Since the free motion Hamiltonian is $H = \mathbf{p}^2/2m = -(1/2m)\nabla^2$, the diffusion equation follows from differentiation with respect to $\hbar/kT$. What is required for quantum statistical mechanics is the trace of the density matrix, that is the integral of the diagonal elements', $\mathbf{r}=\mathbf{r}'$ signifying that the path is closed – a Brownian loop.

The simplicity of free motion allows the path integrals involved in the calculations to be reduced straightaway to ordinary integrals. The reduction is most easily effected directly in words, almost without reference back to the path integral. It seems preferable to collect the relevant path integrals in advance now, and omit them thereafter. They are quoted for Brownian motion, prior to their adaptation from fake time to true time. Brownian motion path integration is due to Wiener [1921]. The first path integral (2), for the probability density just described, is included to set up the notation convention for the other two (3),(4). The quantum particle number density correlation (the result of its evaluation being (9)), is next, with $t_{12}= t_2-t_1$, and the 3×3 quantum particle flux density correlation matrix correlation (the result of its evaluation being (24)) is last. The position $\mathbf{r}$ (of one or both path endpoints) is arbitrary within a (large) container of volume $V$.

$$Q(\mathbf{r}) = \int_{\mathbf{r}(0)=\mathbf{0}}^{\mathbf{r}(t)=\mathbf{r}} \ldots \int \exp\left[-\int_0^t \dot{\mathbf{r}}^2(t')dt'/4D\right] dpath\mathbf{r}(t') = \frac{1}{(4\pi Dt)^{3/2}} \exp[-r^2/4Dt] \quad (2)$$

$$\langle \delta(\mathbf{r}(t_{12})-\mathbf{R})\ \delta(\mathbf{r}(0))\rangle$$
$$= \frac{(4\pi Dt)^{3/2}}{V}\int_{\mathbf{r}(0)=\mathbf{r}}^{\mathbf{r}(t)=\mathbf{r}}\ldots\int \delta(\mathbf{r}(t_2)-\mathbf{r}(t_1)-\mathbf{R})\ \exp\left[-\int_0^t \dot{\mathbf{r}}^2(t')dt'/4D\right] dpath\mathbf{r}(t') \quad (3)$$

$$\langle \dot{\mathbf{r}}(t_{12})\dot{\mathbf{r}}(0)\ \delta(\mathbf{r}(t_{12})-\mathbf{R})\ \delta(\mathbf{r}(0))\rangle$$
$$= \frac{(4\pi Dt)^{3/2}}{V}\int_{\mathbf{r}(0)=\mathbf{r}}^{\mathbf{r}(t)=\mathbf{r}}\ldots\int \delta(\mathbf{r}(t_2)-\mathbf{r}(t_1)-\mathbf{R})\ \dot{\mathbf{r}}(t_2)\ \dot{\mathbf{r}}(t_1)\ \exp\left[-\int_0^t \dot{\mathbf{r}}^2(t')dt'/4D\right] dpath\mathbf{r}(t')$$
$$\quad (4)$$

As mentioned above, the evaluation of correlations and linkings entail true time differences as well as the fake time of reciprocal temperature. The correct procedure for accommodating this is standard but somewhat mysterious. It is loosely summarized as follows; proceed as though all times were fake, and finally analytically continue those times that should be true, by a right angle ('Wick') rotation in the complex plane. For justification, the thorough discussion of Wen [2004] stands out, and the relevant part is extracted in the appendix below. Though perhaps of less interest than the particle flux correlation leading to linking, the particle number density correlation is easier to analyse, and the integrals involved for the former can be evaluated in the same way as for the latter. The next two sections cover the number density correlation, classical and then quantum, and the following two cover the flux density correlation.

*Classical particle number density correlation*

In a classical perfect gas particles move independently in straight lines with a Maxwellian distribution of velocities determined by the temperature *T*. If a particle of mass *m* is known to be at some position at one instant then the probability density for finding it with a vector displacement **R** after time $\tau$ is $(m/2\pi kT\tau^2)^{3/2} \exp[-mR^2/2kT\tau^2]$, where *k* is Boltzmann's constant. It is straightforward, of course, to express this in terms of a particle number density correlation by multiplying by the average particle number density *N/V* for *N* particles in the large volume *V*. To emphasize that correlation in this 'Boltzmann' gas is a single particle effect, *N* is taken as unity. It is easily restored in all formulas henceforth, of course, by replacing *V* by *V/N*.

$$\langle \delta(\mathbf{r}(\tau) - \mathbf{R}) \, \delta(\mathbf{r}(0)) \rangle = \frac{1}{V} \left( \frac{m}{2\pi kT\tau^2} \right)^{3/2} \exp\left[ -\frac{mR^2}{2kT\tau^2} \right] \quad (5)$$

The time integral of this function, though of no apparent significance now, is noted for comparison later. The integral of (5) over all positive times $\tau$ is

$$\frac{1}{VR^2} \sqrt{\frac{m}{8\pi^3 kT}} \quad (6)$$

That very brief account concludes this section.

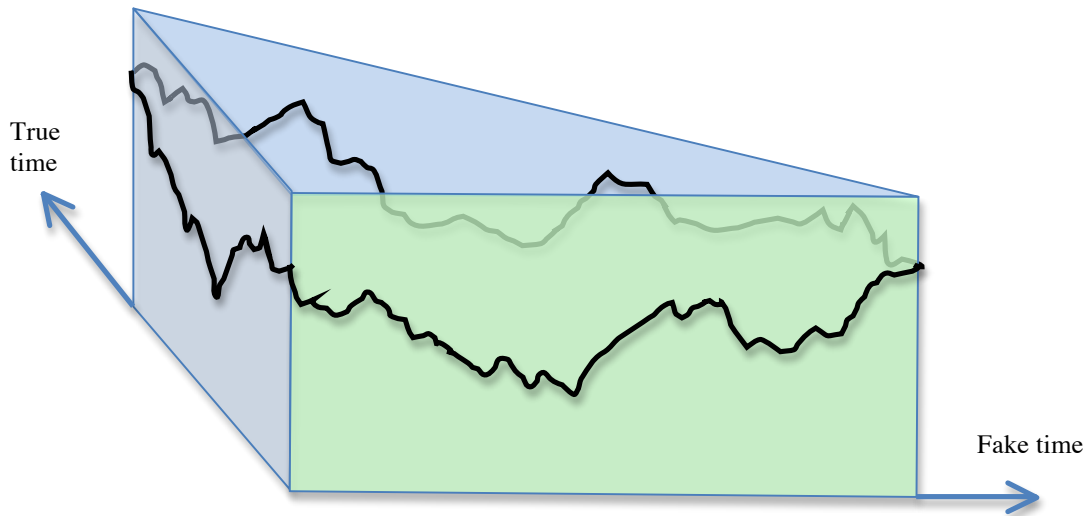

Fig 1. A Brownian motion in fake time (reciprocal temperature) is shown schematically in the front panel. A single space coordinate is drawn vertical and the end values are equal - it ends where it started - a Brownian loop. This would suffice for the quantum statistical mechanics of a free particle provided (true) time is not involved. Otherwise a detour is required, shown as the back route. One portion of this, representing purely Feynman motion, takes place at constant fake time. Such a detour is needed for the present (true) time correlation calculations, but it can be accessed analytically from the purely Brownian front panel.

*Quantum particle number density correlation*

In the quantum kinetic theory of a perfect gas, the straight line classical motion is replaced by an infinity of jittery Feynman motions. But appended to this, indeed substituting for it analytically, is the jittery Brownian motion in fake time (reciprocal temperature). This serves, one might say, to 'bring the temperature down' from infinity to the finite desired temperature. The situation is pictured in Fig 1 where the horizontal axes are true and fake time and the vertical one represents space (with two dimensions suppressed). The present correlation quests involve true time (as well as fake time), and therefore require the detour via the 'back route'. However the calculation can be confined to the simpler 'front route' which corresponds to asking for the same correlations in the context of Brownian motion enacting the diffusion equation, rather than Feynman motion enacting the Schrödinger equation. Finally the detour can be accommodated by analytic extension with a standard prescription.

The diffusive spreading of an initial point ($\delta$-function) is described (2) by the 3-dimensional Gaussian $(4\pi Dt)^{-3/2} \exp[-r^2/4Dt]$, where $t$ is the duration and $D$ is the diffusion coefficient (*Flux density =D × Number density gradient*). It will be useful to think of the exponent of the Gaussian as ($-1/4D$ times) a hypothetical 'velocity squared times duration', $r^2/t^2 \times t$. Because the thermal quantum problem involves the trace of operators (the Boltzmann weighted sum over eigenstates, as in the appendix) the relevant specialization of Brownian motion is to motions that end up, after time *t*, where they started – a loop. Brownian loops implement the trace. The task in this section is to illustrate the calculation process in a more routine instance: that of the distribution of chord vectors of a Brownian loop.

For a Brownian loop $\mathbf{r}(\tau)$ of duration *t* (having $\mathbf{r}(0)=\mathbf{r}(t)=\mathbf{r}$), the chord vectors are defined by $\mathbf{R}=\mathbf{r}(t_2)-\mathbf{r}(t_1)$ for all $0< t_1< t_2<t$. These chord vectors have a probability density $P(\mathbf{R}|t_1,t_2)$ which is sought, where this is the integral (7) of the probability density, given a starting and finishing point at $\mathbf{r}$, that the position $\mathbf{r}_2$ at time $t_2$, differs by the vector $\mathbf{R}$ from that, $\mathbf{r}_1$, at time $t_1$. Because the path is a loop, the distribution of chords is dependent only on the duration $t_{12}= t_2-t_1$ and not on $t_2$ and $t_1$ individually: $P(\mathbf{R}|t_1,t_2)= P(\mathbf{R}|t_{12})$. The dramatic simplicity of free space path integrals is that they evaluate exactly in terms of the classical motions from one event to the next; that is, in terms of piecewise constant velocity sections: from $\mathbf{r}$ to $\mathbf{r}_1$ to $\mathbf{r}_2$ and back to $\mathbf{r}$, a spatial triangle. With $t_{21}= t-t_{12}$ the Gaussian integrals over all $\mathbf{r}_1$ and $\mathbf{r}_2$ render the result independent of $\mathbf{r}$:

$$P(\mathbf{R}|t_1,t_2) = P(\mathbf{R}|t_{12}) = (4\pi Dt)^{3/2} \int d^3\mathbf{r}_1 \int d^3\mathbf{r}_2 (4^3\pi^3 D^3 t_1 t_{12}(t-t_2))^{-3/2} \times$$

$$\times \delta(\mathbf{R}-\mathbf{r}_2+\mathbf{r}_1) \exp\left(-\frac{(\mathbf{r}_1-\mathbf{r})^2}{4Dt_1} - \frac{(\mathbf{r}_2-\mathbf{r}_1)^2}{4Dt_{12}} - \frac{(\mathbf{r}-\mathbf{r}_2)^2}{4D(t-t_2)}\right) \quad (7)$$

$$= \left(\frac{t}{4\pi D t_{12} t_{21}}\right)^{3/2} \exp\left[-\frac{\mathbf{R}^2}{4D}\left(\frac{1}{t_{12}}+\frac{1}{t_{21}}\right)\right] \quad (8)$$

The prefactor $(4\pi Dt)^{3/2}$ is the reciprocal of the same double integral except without the δ-function, ensuring that the $P(\mathbf{R}|t_1,t_2)$ is correctly normalized to unity. The resulting expression (8) is neatly interpreted in terms of a collapsed version of the triangle, a line segment from $\mathbf{r}$ to $\mathbf{r}+\mathbf{R}$ and back to $\mathbf{r}$. The exponent is proportional to the total of 'velocity squared times duration' of the legs of this collapsed there-and-back excursion: the outward journey time being $t_{12}$, and the return time $t_{21}$. The number density correlation of this Brownian loop motion is just $P(\mathbf{R}|t_{12})$ divided by the (large) container volume *V* to which $\mathbf{r}$ is confined, as in (5). Since there is no dependence on $\mathbf{r}$, it may be taken as the origin

$$\langle \delta(\mathbf{r}(t_{12})-\mathbf{R}) \, \delta(\mathbf{r}(0)) \rangle = \frac{1}{V} P(\mathbf{R}|t_{12}) \quad (9)$$

Before moving on to the quantum mechanics, the sequence of equalities in (10) and (11) below represent a brief detour highlighting that there can (and will) be interest in integrating such correlation functions over the full range of time (this appears on the left in equation (11)). The outcome, on the left of equation (10) is the marginal probability $P(\mathbf{R})$ of chord vector $\mathbf{R}$ for a Brownian loop of duration *t*, irrespective of the chord duration $t_{12}$:

$$P(\mathbf{R}) = \frac{\int_0^t dt_2 \int_0^{t_2} dt_1 P(\mathbf{R}|t_1,t_2)}{\int_0^t dt_2 \int_0^{t_2} dt_1} = \int_0^t P(\mathbf{R}|t_{12}) \frac{2(t-t_{12})}{t^2} dt_{12} \quad (10)$$

$$= \int_0^t P(\mathbf{R}|t_{12}) \frac{1}{t} dt_{12} = \frac{1}{(2\pi Dt/3)^{3/2}} \exp\left(-\frac{3\mathbf{R}^2}{2Dt}\right) \quad (11)$$

The integrand in (11) has been simplified from that in (10) by the fact that $P(\mathbf{R}|t_{12})$ is symmetrical about $t_{12} = \tfrac{1}{2}t$ meaning that the integral of $P(\mathbf{R}|t_{12}) \times (t_{12} - \tfrac{1}{2}t)$ is zero. Substituting (8) into (11), the integral over $t_{12}$ can be performed by centring the time $(t_{12} - \tfrac{1}{2}t)$ and then using the substitution that renders the exponential a simple decay (or by referring, for example, to table entry A5 of [Feynman and Hibbs 1965]). Similar integrals will arise for the quantum linking problem later.

This Brownian analysis now needs adapting to thermal Quantum mechanics. The correlation corresponding to (9) is simply supplied, as indicated in the introduction, by the formal procedure of taking $t_{12}$ to be $i\tau$ and $t_{21}$ to be $t - i\tau$. The quantum particle number density correlation is, then

$$\langle \delta(\mathbf{r}(\tau) - \mathbf{R}) \, \delta(\mathbf{r}(0)) \rangle = \frac{1}{V} P(\mathbf{R}|i\tau)$$

$$= \frac{1}{V} \left(\frac{t}{4\pi Di\tau(t-i\tau)}\right)^{3/2} \exp\left(-\frac{\mathbf{R}^2 t}{4Di\tau(t-i\tau)}\right) \quad (12)$$

Here the thermal average angle brackets on the left hand side require an updated definition for quantum mechanics, and this is supplied in the appendix, following Wen [2004]. The classical limit $\hbar \to 0$ of (12) is (5) with the substitutions (1), $D = \hbar/2m$ and $t = \hbar/kT$.

The time integral of (12) is now to be evaluated. Like (5) in the classical case, there is little physical motivation for this, but the calculation exemplifies, in a simpler context, similar ones that have significance in the particle flux correlation of the next two sections. The integral over the positive real axis of $\tau$ can be split into its real and imaginary parts by deforming the contour to have a right angle bend: $0 \to -\tfrac{1}{2}it \to \infty - \tfrac{1}{2}it$. Adjusting the integration variable to be real and to start from zero separately in each stretch one has:

$$\frac{1}{V} \int_0^\infty d\tau \left(\frac{t}{4\pi Di\tau(t-i\tau)}\right)^{3/2} \exp\left(-\frac{\mathbf{R}^2 t}{4Di\tau(t-i\tau)}\right) \quad (13)$$

$$= \frac{1}{V}\int_0^\infty d\tau \left(\frac{t}{4\pi D((\tfrac{1}{2}t)^2 + \tau^2)}\right)^{3/2} \exp\left(-\frac{\mathbf{R}^2 t}{4D((\tfrac{1}{2}t)^2 + \tau^2)}\right)$$

$$-i\frac{1}{V}\int_0^{\tfrac{1}{2}t} d\tau \left(\frac{t}{4\pi D((\tfrac{1}{2}t)^2 - \tau^2)}\right)^{3/2} \exp\left(-\frac{\mathbf{R}^2 t}{4D((\tfrac{1}{2}t)^2 - \tau^2)}\right)$$
(14)

$$= \frac{1}{4\pi VRD}\exp\left(-\frac{\mathbf{R}^2}{Dt}\right)\left[\mathrm{erfi}\left(\sqrt{\frac{\mathbf{R}^2}{Dt}}\right) - i\right] \tag{15}$$

The two integrals in (14) differ only in the signs in front of each $\tau^2$ in the integrands, and the limits on the integrals. They can each be evaluated by the substitution that renders the exponential a simple decay. Since $\exp[-x^2]\,\mathrm{erfi}[x] \sim x/\sqrt{\pi}$ as $x$ tends to $\infty$, the classical limit $\hbar \to 0$ reduces to (6)

$$\frac{1}{4\pi VRD}\sqrt{\frac{Dt}{R^2}}\frac{1}{\sqrt{\pi}} = \frac{1}{VR^2}\sqrt{\frac{m}{8\pi^3 kT}} \tag{16}$$

using the substitutions (1).

*Classical particle flux density correlation*

Here the relevant quantity in kinetic theory is the average frequency with which a particle on its straight line path passes through a fixed infinitesimal area element $\mathbf{\Delta A}_1$ and then through another $\mathbf{\Delta A}_2$ located at displacement $\mathbf{R}$ away, between times $\tau$ and $\tau+d\tau$ later. (An equal average frequency applies with subscripts 1 and 2 exchanged). Of course the average frequency is infinitesimal in proportion to the product $|\mathbf{\Delta A}_1|\,|\mathbf{\Delta A}_2|\,d\tau$. It factorizes into a geometrical part, the frequency with which its straight-line path passes through both elements irrespective of the value of delay, and the delay probability distribution.

The delay distribution

$$\frac{m^2 R^4}{2k^2 T^2}\frac{1}{\tau^5}\exp\left(-\tfrac{1}{2}\frac{mR^2}{kT\tau^2}\right) \tag{17}$$

derives from the Maxwellian distribution of speeds $v$ at the moment of passage, $(m^2/2k^2T^2)\,v^3\exp(-\tfrac{1}{2}mv^2/kT)\,dv$, with the relation $v=R/\tau$, and therefore $|dv/d\tau|=R/\tau^2$. It integrates over all positive $\tau$ to unity.

The geometrical part is itself the product of several factors; it also arises in the theory of thermal radiative exchange, as noted later.

$$\left\{\frac{1}{4V}\sqrt{\frac{8kT}{\pi m}}|\Delta\mathbf{A}_1|\right\}\left[\frac{1}{\pi}\frac{\mathbf{R}}{R}\cdot\frac{\Delta\mathbf{A}_1}{|\Delta\mathbf{A}_1|}\right]\left(\frac{\mathbf{R}\cdot\Delta\mathbf{A}_2}{R^3}\right) \tag{18}$$

$$= \frac{1}{VR^4}\sqrt{\frac{kT}{2\pi^3 m}}(\mathbf{R}\cdot\Delta\mathbf{A}_1)(\mathbf{R}\cdot\Delta\mathbf{A}_2) \tag{19}$$

The round bracket in (18) is the solid angle subtended by $\Delta\mathbf{A}_2$ at the points of $\Delta\mathbf{A}_1$. The square bracket is the probability per unit solid angle for random straight lines through $\Delta\mathbf{A}_1$ to lie in the direction of $\mathbf{R}$. (It integrates to unity over $2\pi$ of solid angle). The brace bracket is the rate at which a particle in a volume $V$ passes through $\Delta\mathbf{A}_1$ (just one way through, not both). The same result would apply for the reversed set of trajectories passing through $\Delta\mathbf{A}_2$ first, then $\Delta\mathbf{A}_1$. The classical particle flux correlation, then, is the average brackets in (20), the 3×3 tensor such that:

$$\Delta\mathbf{A}_2\cdot\langle\mathbf{j}(\mathbf{R},\tau)\,\mathbf{j}(0,0)\rangle\cdot\Delta\mathbf{A}_1$$
$$= \sqrt{\frac{kT}{2\pi m}}\frac{(\mathbf{R}\cdot\Delta\mathbf{A}_1)(\mathbf{R}\cdot\Delta\mathbf{A}_2)}{\pi VR^4}\frac{m^2R^4}{2k^2T^2}\frac{1}{\tau^5}\exp\left(-\frac{1}{2}\frac{mR^2}{kT\tau^2}\right) \tag{20}$$

where the right hand side is the product of (19) and (17). The time integral of (20) over the positive real axis of $\tau$ yields (19) since the time integral of (17) is unity.

One can extend the geometrical part (19) from its two infinitesimal areas to two finite size pieces of surface of arbitrary shape and location. Adapting the result of Sparrow [1963] for radiative exchange (as a shortcut to a more general discussion in (32) later), the average rate at which a particle passes through one surface and then the other (with signs for the senses of passage) is given by the following double boundary integral. (This is valid however convoluted the surfaces may be; the radiative exchange circumstance has more restricted validity, namely all pairs of surface points must be mutually visible, fixing the signs of passage, otherwise more sophisticated considerations apply [Hottle 1954][Derrick 1985]).

$$\frac{1}{V}\sqrt{\frac{kT}{8\pi^3 m}}\oint_1\oint_2 \log|\mathbf{r}_1-\mathbf{r}_2|\, d\mathbf{r}_1\cdot d\mathbf{r}_2 \tag{21}$$

(The opposite order gives an equal rate)

*Quantum particle flux density correlation*

What is required is similar to that required for the quantum particle number density correlation earlier. Once again the procedure is to perform the calculation for the Brownian particle flux density correlation, and then rotate appropriately in the complex plane for the quantum particle flux correlation.

Now, for flux density, the chord probability for the classically for the Brownian loop is weighted with the product of velocities at each of the chord endpoint events 1 and 2.

To incorporate these (exactly), one can split each of the two events into a pair infinitesimally apart in time by $\Delta t$, so the 'return trip' becomes a 3D quadrilateral in space. (Just as the there-and back excursion earlier in the number density correlation, arose from the collapse of a triangle, so here the quadrilateral arises from the collapse of a pentagon). Some care is needed not to bias the infinitesimal sections in favour of one or other of the two finite sections (operator-wise this would correspond to inappropriate operator ordering). Various unbiased strategies are possible leading to the same (correct) result. The easiest is adopted, and is described in the next paragraph. (Another would be to consider the four different quadrilateral types in which a pair of vertices lie at the ends of the fixed dashed line (rather than the dashed line ends being midpoints), and then average over the four types.

The quadrilateral should be pictured as long and thin (Fig 2), although formally the two new edge vectors $\mathbf{\Delta R}_1$ and $\mathbf{\Delta R}_2$, which represent the two velocities (times $\Delta t$), are each to be integrated over all space.

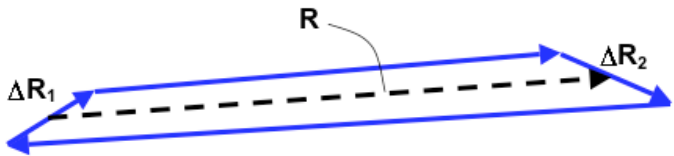

Fig 2. A quadrilateral loop (solid line, typically not planar) in 3D used for the calculation of the correlation of the particle flux density at the two ends of the given dashed vector $\mathbf{R}$. The quadrilateral is such that the ends of $\mathbf{R}$ lie at the midpoints of two opposite legs $\Delta \mathbf{R}_1$ and $\Delta \mathbf{R}_2$ but is otherwise arbitrary. Indeed $\Delta \mathbf{R}_1$ and $\Delta \mathbf{R}_2$ are each to be integrated over all directions and magnitudes. The quadrilateral is the locus of a hypothetical piecewise constant velocity motion, the legs having durations: $t_{12}$, $\Delta t$, $t_{21}$, $\Delta t$, with $\Delta t$ being infinitesimal so the two velocities being correlated are $\Delta \mathbf{R}_1 / \Delta t$ and $\Delta \mathbf{R}_2 / \Delta t$. The vector $\mathbf{R}$ represents the separation vector between two infinitesimal test circuits with arbitrary orientation (not shown).

In full, the four leg displacement vectors are $\mathbf{R}-½\mathbf{\Delta R}_1-½\mathbf{\Delta R}_2$, $\mathbf{\Delta R}_1$, $-\mathbf{R}-½\mathbf{\Delta R}_1-½\mathbf{\Delta R}_2$, $\mathbf{\Delta R}_2$ with corresponding durations $t_{12}=t_2-t_1$, $\Delta t$, $t_{21}=t-t_{12}$, $\Delta t$, with $\Delta t$ infinitesimal. The velocity-product-weighted-integral supplying the product of the particle fluxes through the infinitesimal circuits is then dependent only on the durations and not their starting time $t_1$ (as in (7) earlier). It is given by

$$\Delta \mathbf{A}_2 \cdot \langle \mathbf{j}(\mathbf{R}, t_{12}) \, \mathbf{j}(\mathbf{0},0) \rangle \cdot \Delta \mathbf{A}_1 = \langle (\dot{\mathbf{r}}(t_{12}) \cdot \Delta \mathbf{A}_2)(\dot{\mathbf{r}}(0) \cdot \Delta \mathbf{A}_1) \, \delta(\mathbf{r}(t_{12}) - \mathbf{R}) \, \delta(\mathbf{r}(0)) \rangle$$

$$= \frac{(4\pi Dt)^{3/2}}{V} \iint \frac{\Delta \mathbf{R}_1 \cdot \Delta \mathbf{A}_1}{\Delta t} \frac{\Delta \mathbf{R}_2 \cdot \Delta \mathbf{A}_2}{\Delta t} \frac{\exp(-S) \, d^3\Delta \mathbf{R}_1 d^3\Delta \mathbf{R}_2}{(4^4 D^4 \pi^4 t_{12} t_{21} \Delta t^2)^{3/2}} \quad (22)$$

where the 'action' $S$, proportional to the total of 'velocity squared times duration' around the four legs of the quadrilateral, is

$$S(\Delta \mathbf{R}_1, \Delta \mathbf{R}_2) = \frac{(\mathbf{R} - \tfrac{1}{2}\Delta \mathbf{R}_1 - \tfrac{1}{2}\Delta \mathbf{R}_2)^2}{4Dt_{12}} + \frac{\Delta \mathbf{R}_1^2}{4D\Delta t} + \frac{(-\mathbf{R} - \tfrac{1}{2}\Delta \mathbf{R}_1 - \tfrac{1}{2}\Delta \mathbf{R}_2)^2}{4Dt_{21}} + \frac{\Delta \mathbf{R}_2^2}{4D\Delta t} \quad (23)$$

As usual with path expressions, that between the two equalities in (22) is operator-free, and thus not subject to ordering concerns (instead the quadrilateral geometry chosen encodes this subtlety, as mentioned). Having a quadratic action, the spatial integrations in (22) are mathematically routine, but capable of some interpretation, given after (24). The factors are left unsimplified to accord with this interpretation, in particular the $\Delta t$ will cancel out as it should. So (22) with (23) yields

$$\Delta \mathbf{A}_2 \cdot \langle \mathbf{j}(\mathbf{R}, t_{12}) \, \mathbf{j}(\mathbf{0},0) \rangle \cdot \Delta \mathbf{A}_1 = \frac{(4\pi Dt)^{3/2}}{V} \frac{(4\pi D\Delta t)^3}{(4^4 D^4 \pi^4 t_{12} t_{21} \Delta t^2)^{3/2}} \times \exp\left(-\frac{R^2 t}{4D t_{12} t_{21}}\right)$$

$$\times \left[ (\mathbf{R} \cdot \Delta \mathbf{A}_1)(\mathbf{R} \cdot \Delta \mathbf{A}_2) \frac{1}{4}\left(\frac{t_{12} - t_{21}}{t_{12} t_{21}}\right)^2 - (\Delta \mathbf{A}_1 \cdot \Delta \mathbf{A}_2) 2D \frac{1}{4}\left(\frac{t}{t_{12} t_{21}}\right) \right] \quad (24)$$

As a one-paragraph digression it may be helpful to interpret how the terms in this expression arise dynamically. The right hand side of (22) can be thought of as *Integral of velocity product=Average of velocity product × Integral without velocity product*. For the latter one has

$$\int \exp(-S) d^3\Delta \mathbf{R}_1 d^3\Delta \mathbf{R}_2 = (4\pi D\Delta t)^3 \exp\left(-\frac{R^2 t}{4D t_{12} t_{21}}\right) \quad (25)$$

Here the exponent on the right is the action of the stationary action loop $\partial S / \partial \Delta \mathbf{R}_1 = \partial S / \partial \Delta \mathbf{R}_2 = 0$. This loop has $\Delta \mathbf{R}_1 = \Delta \mathbf{R}_2$ collinear along $\pm \mathbf{R}$, with magnitude $\tfrac{1}{2} R\Delta t \, |t_{12}^{-1} - t_{21}^{-1}| = \tfrac{1}{2} R\Delta t \, |t_{12} - t_{21}| / t_{12} t_{21}$ (using $\Delta t$ infinitesimal). The prefactor in (25) is $\sqrt{(2\pi)^6 / \det S''}$, where $S''$ is the 6×6 matrix of second derivatives of $S$. Since the coordinates separate, this is block diagonal with three identical 2x2 blocks. Its determinant is dominated, since $\Delta t$ is infinitesimal, by the by the diagonal product $1/(2D\Delta t)^6$. For the average of the velocity product in (22) there are two contributions, the stationary action product supplies the first term in the square bracket in (22), and fluctuations of the $\Delta \mathbf{R}$s about this stationary action path supply the second term. This latter comes from the

(equal) non-zero off-diagonal elements of the inverse of the matrix $S''$. These elements are $-((t_{12}^{-1} + t_{21}^{-1})/8D)/(1/2D\Delta t)^2$. Separate directions $x,y,z$ are uncoupled, so this multiplies $\Delta A_1 \cdot \Delta A_2$.

So, returning to (24) with the substitutions $i\tau$ for $t_{12}$ and $t - i\tau$ for $t_{21}$ in (24), the result for the quantum flux density correlation is

$$\Delta \mathbf{A}_2 \cdot \langle \mathbf{j}(\mathbf{R},\tau)\, \mathbf{j}(0,0) \rangle \cdot \Delta \mathbf{A}_1 = \frac{1}{V}\sqrt{\frac{t^3}{4^5 \pi^3 D^3}} \left\{ \frac{1}{(i\tau(t-i\tau))^{7/2}} \right\} \exp\left(-\frac{R^2 t}{4Di\tau(t-i\tau)}\right) \times$$
$$\times \left[ (\mathbf{R} \cdot \Delta \mathbf{A}_1)\,(\mathbf{R} \cdot \Delta \mathbf{A}_2)\,(t^2 - 4i\tau(t-i\tau)) - (\Delta \mathbf{A}_1 \cdot \Delta \mathbf{A}_2)\, 2Di\tau(t-i\tau)t \right] \quad (26)$$

There is an ambiguity of overall sign arising from the 7/2 power of a complex number in the brace bracket. The correct choice of root is that for which the real part of the brace bracket is negative. One way of determining this is from the classical limit $\hbar \to 0$. With $D = \hbar/2m$ and $t = \hbar/kT$ from (1), this picks out the term $-(\mathbf{R} \cdot \Delta \mathbf{A}_1)(\mathbf{R} \cdot \Delta \mathbf{A}_2)4\tau^2$ in the square bracket, so the stated choice is needed for agreement with the positive classical result (20). (This choice is implemented later with modulus signs in (28), yielding (29) for the all-time integral of (26)).

In a sense this is already topological since passage through an infinitesimal area, or its infinitesimal boundary hoop, is a definite event, it has definite timing incrementing the passage number by ±1. To extend this, though, to passage through finite size hoops, the timing is an obstacle – it depends on the surfaces chosen to span the boundary hoops and register the passages. For the purpose of obtaining a properly topological quantity that depends only on the hoops (in the next section), the timing first needs to be eliminated by all-time integration of the flux density correlation. Afterwards Stokes' theorem (doubly applied) can be used to effect the area integrations, converting the double surface integral to a double boundary line (hoop) integration.

The integral of the quantum flux correlation (26) over all positive delay times $\tau$ can be performed by first rearranging it to the form of the integrand in (27). Then, by deformation of the contour of integration to $0 \to -\tfrac{1}{2}it \to \infty - \tfrac{1}{2}it$ (just as (13) was deformed to (14)), the integrals can be evaluated in the same manner as (14) to give (29).

$$\int_0^\infty \Delta \mathbf{A}_2 \cdot \langle \mathbf{j}(\mathbf{R},\tau)\, \mathbf{j}(0,0) \rangle \cdot \Delta \mathbf{A}_1\, d\tau$$

$$= \int_0^\infty d\tau\, \frac{1}{V}\sqrt{\frac{t^3}{4^5 \pi^3 D^3}} \exp\left(-\frac{R^2 t}{4Di\tau(t-i\tau)}\right) \times$$
$$\left[ \frac{(\mathbf{R} \cdot \Delta \mathbf{A}_1)(\mathbf{R} \cdot \Delta \mathbf{A}_2)\, t^2}{(i\tau(t-i\tau))^{7/2}} - \frac{4(\mathbf{R} \cdot \Delta \mathbf{A}_1)(\mathbf{R} \cdot \Delta \mathbf{A}_2) + (\Delta \mathbf{A}_1 \cdot \Delta \mathbf{A}_2)2Dt}{(i\tau(t-i\tau))^{5/2}} \right] \quad (27)$$

$$= \frac{1}{V}\sqrt{\frac{t^3}{4^5\pi^3 D^3}} \int_0^\infty d\tau \exp\left(-\frac{R^2 t}{4D((\frac{1}{2}t)^2+\tau^2)}\right) \times$$

$$\times \left[\frac{(\mathbf{R}\cdot\Delta\mathbf{A}_1)(\mathbf{R}\cdot\Delta\mathbf{A}_2)t^2}{-\left|((\frac{1}{2}t)^2+\tau^2)^{7/2}\right|} - \frac{4(\mathbf{R}\cdot\Delta\mathbf{A}_1)(\mathbf{R}\cdot\Delta\mathbf{A}_2)+(\Delta\mathbf{A}_1\cdot\Delta\mathbf{A}_2)2Dt}{-\left|((\frac{1}{2}t)^2+\tau^2)^{5/2}\right|}\right] \quad (28)$$

$$-i\frac{1}{V}\sqrt{\frac{t^3}{4^5\pi^3 D^3}} \int_0^{\frac{1}{2}t} d\tau \left(\begin{array}{c}\text{Same integrand except that each } \tau^2 \\ \text{has a minus sign in front of it}\end{array}\right)$$

$$= (\mathbf{R}\cdot\Delta\mathbf{A}_1)(\mathbf{R}\cdot\Delta\mathbf{A}_2) \times \frac{1}{V}\sqrt{\frac{t^3}{4^5\pi^3 D^3}} \times$$

$$\frac{4}{R^5 t^3}\left(12D^2 Rt - 2D\sqrt{\pi Dt}(2R^2+3Dt)\exp\left(-\frac{R^2}{Dt}\right)\left(\operatorname{erfi}\left(\sqrt{\frac{R^2}{Dt}}\right)-i\right)\right)$$

$$+ \Delta\mathbf{A}_1\cdot\Delta\mathbf{A}_2 \times \frac{1}{V}\sqrt{\frac{t^3}{4^5\pi^3 D^3}} \times \quad (29)$$

$$\frac{4}{R^3 t^3}\left(-4D^2 Rt + 2D\sqrt{\pi Dt}(2R^2+Dt)\exp\left(-\frac{R^2}{Dt}\right)\left(\operatorname{erfi}\left(\sqrt{\frac{R^2}{Dt}}\right)-i\right)\right)$$

The modulus signs in (28) disambiguate the overall sign in the manner described after (26). The classical limit $\hbar \to 0$ (with $D=\hbar/2m$ and $t=\hbar/kT$ from (1)) is real, obtained by using $\exp(-s^2)\operatorname{erfi}(s) \to 1/(\sqrt{\pi}\,s)$ as $s$ tends to infinity. (This means that in the large bracket coefficient of $\Delta\mathbf{A}_1\cdot\Delta\mathbf{A}_2$, the first two terms dominate, but cancel, leaving only the dependence on $(\mathbf{R}\cdot\Delta\mathbf{A}_1)(\mathbf{R}\cdot\Delta\mathbf{A}_2)$). Then the limit of (29) is correctly (19), the integral of (17) being unity.

*Linking consequence*

The average product of the two linking numbers of the particle's path, with two arbitrary fixed hypothetical hoops, increases linearly in time. As a topological feature (functional) of the two hoops in a perfect gas, this rate of increase invites calculation. It is to be supplied by using Stokes' theorem to convert the flux density to a flux, or more precisely, the time integrated correlation of the former to that of the latter. This equals half the desired rate of increase of the average; the 'half' arises from the fact that the linking number product also includes paths that go through the hoops in reverse order; only one order has been accounted for in the flux.

Some physical motivation for the evaluation of this quantity might come from electromagnetism. If the particle were charged with charge $q$, then its passage completely through a hoop would induce a pulse of magnetic circulation with time-integrated value $(q\mu_0)^{-1}\int_{-\infty}^{\infty} dt \oint \mathbf{B}(\mathbf{r},t)\cdot d\mathbf{r}$. The instantaneous circulation is not accessible analytically from the charge flux (electric current) because of the electric field term in the Maxwell equation $Curl \mathbf{B} = \mu_0 \mathbf{j} + \varepsilon_0 \mu_0 \partial \mathbf{E}/\partial \tau$, but the all-time integration eliminates this electric contribution. So the rate of increase of the average linking number product supplies the thermal average

$$(1/q^2\mu_0^2)\,(1/2t)\int_{-t}^{t} dt_1 \left(\oint_1 \mathbf{B}(\mathbf{r}_1,t_1)\cdot d\mathbf{r}_1\right)\ \int_{-t}^{t} dt_2 \left(\oint_2 \mathbf{B}(\mathbf{r}_2,t_2)\cdot d\mathbf{r}_2\right)$$
$$\xrightarrow[t\to\infty]{}\ (1/q^2\mu_0^2)\int_{-\infty}^{\infty} d\tau \left\langle \left(\oint_1 \mathbf{B}(\mathbf{r}_1,\tau)\cdot d\mathbf{r}_1\right)\left(\oint_2 \mathbf{B}(\mathbf{r}_2,0)\cdot d\mathbf{r}_2\right)\right\rangle \quad (30)$$

The gas of particles, if there is more than one particle, must be assumed perfect, non-interacting, despite their charge. It may be that this magnetic consequence of thermal Feynman motion comes closest to being a measurable manifestation of particle flux correlation, classical or quantum. Perhaps thin hoop shaped solenoids (conducting helices) could detect magnetic circulation along their hoop axes. In a vacuum with no charged particles, the magnetic circulation of thermal electromagnetic fields is zero around any hypothetical hoop, of course, by Maxwell's equations, and so do not contribute.

The double-area to double-line use of Stokes' theorem arises in a couple of simpler well-known physical contexts: thermal 'radiative transfer' between two 'black' surfaces, and 'mutual inductance' between two conducting circuits. Those two, and the present context, are all flux-type spatial connections (radiative heat flux, magnetic flux, particle flux) and, all being divergence-free fluxes, the isotropy of space guarantees that they have the simple form

$$\oint_1 \oint_2 f(|\mathbf{r}_2-\mathbf{r}_1|)\, d\mathbf{r}_1\cdot d\mathbf{r}_2 \quad (31)$$

with different functions $f$ in the three contexts. In fact both the two simpler contexts are exhibited in different limits of the present time integrated particle flux correlation. These are mentioned after the result (34).

To write the relationship of the double line integral to the double area integral, it is useful to express $f$ itself as an (indefinite) integral, $f(R) = \int_R^{const} \tilde{R}\lambda(\tilde{R})d\tilde{R}$ with $\mathbf{R} = |\mathbf{R}|\hat{\mathbf{R}} = \mathbf{r}_2 - \mathbf{r}_1$ and $\mathbf{n}_1$ and $\mathbf{n}_2$ being unit normals to the area elements (the value of 'const' is immaterial since $\oint d\mathbf{r} = 0$). Then [Hannay 2019], one has

$$\oint_1\oint_2 f(R)d\mathbf{r}_1\cdot d\mathbf{r}_2 = \oint_1\oint_2\left(\int_R^{const}\tilde{R}\lambda(\tilde{R})d\tilde{R}\right)d\mathbf{r}_1\cdot d\mathbf{r}_2$$
$$= \int d^2\mathbf{r}_1 \int d^2\mathbf{r}_2 \left[(\mathbf{n}_1\cdot\mathbf{n}_2)(R\lambda'+2\lambda) - R\lambda'(\mathbf{n}_1\cdot\hat{\mathbf{R}})(\mathbf{n}_2\cdot\hat{\mathbf{R}})\right] \quad (32)$$

For radiative exchange the function $\lambda(\tilde{R})$ has $\lambda \propto 1/\tilde{R}^2$ so $f \propto \log(R)$, while for mutual inductance $\lambda \propto 1/\tilde{R}^3$ so $f \propto 1/R$. For the present problem, one can notice that, conveniently in (32), $\frac{1}{2}\left[(\tilde{R}\lambda'+2\lambda) - \tilde{R}\lambda'\right] = \lambda$, so $\lambda(\tilde{R})$ is half the sum of the coefficients in (29) of $(\mathbf{R}\cdot\Delta\mathbf{A}_1)(\mathbf{R}\cdot\Delta\mathbf{A}_2)/R^2$ and $\Delta\mathbf{A}_1\cdot\Delta\mathbf{A}_2$.

$$\lambda(\tilde{R}) = \frac{1}{4V\tilde{R}^3}\sqrt{\frac{D}{\pi^3 t}}\left[2\tilde{R} - \sqrt{\pi Dt}\exp\left(-\frac{\tilde{R}^2}{Dt}\right)\left(\text{erfi}\left(\frac{\tilde{R}}{\sqrt{Dt}}\right) - i\right)\right] \tag{33}$$

In the integral (31) for $f(R)$, any value of the limit 'const' can be chosen, as remarked, and natural choices are zero or infinity to avoid apparent arbitrariness. Awkwardly, the real part of the integral fails to converge at infinity, and the imaginary part fails to converge at zero. However there is no objection to separate choices of limit for each part, thus one may take $f$ as follows and evaluate the integrals. With $_2F_2$ as the generalized hypergeometric function

$$f(R) \equiv \int_R^0 \tilde{R}\ \text{Re}[\lambda(\tilde{R})]\ d\tilde{R} + i\int_R^\infty \tilde{R}\ \text{Im}[\lambda(\tilde{R})]\ d\tilde{R}$$

$$= \frac{R^2}{6V}\sqrt{\frac{1}{\pi^3 t^3 D}}\ _2F_2\left[\{1,1\}, \{2,\tfrac{5}{2}\}, -\frac{R^2}{Dt}\right]$$

$$+ i\frac{D}{4V}\sqrt{\frac{1}{\pi^3}}\left(\sqrt{\frac{\pi}{Dt}}\text{erfc}\left(\frac{R}{\sqrt{Dt}}\right) - \frac{1}{R}\exp\left(-\frac{R^2}{Dt}\right)\right)$$

$$\tag{34}$$

This is the final result; substituted into (32) it gives the particle flux correlation through the two hoops integrated over all positive time. The motivation for calculating this was that it had topological significance, being dependent only on the hoops, independent of spanning surfaces. This was phrased in terms of a rate of growth of the average linking number product of the random path with the two hoops. The result (34) however is a complex number - quantum statistical correlations are entitled to be so. A geometrical picture to reinforce the linking interpretation of the real part (the $_2F_2$ part) does not seem obvious, though (34) with (31) correctly limits to (21) in the classical limit $\hbar \to 0$ (invoking $\oint d\mathbf{r} = 0$).

A linking interpretation of the imaginary part is more definite, and quite striking (though perhaps the definiteness is to be expected since in general Hermitian contexts the imaginary part of a quantum correlation is a physical response function [Wen 2004]). The imaginary part of (32), via (33) and (34), equals $1/t$ ×1/2× the average product of the two linking numbers of a Brownian loop of duration $t$ (having diffusion coefficient $D = \hbar/2m$) with the two hoops. This average linking number product was obtained earlier by a different method (based on work on topology in optics) [Hannay 2019, 2018, 2001, 1995]. The low temperature (i.e. large $t$) limit of (34) comes from the imaginary part ($f \to -i\ D/(4\pi^{3/2}VR)$) and thus yields (a constant times) the mutual inductance double line integral [Feynman et al 1964] of the two hoops considered as electrical conductors [Hannay 2019]. Also treated in Hannay [2019] is the related limit in which the two hoops nearly coincide to become a single hoop. Like mutual inductance, the average Brownian linking number product diverges in the coincidence limit showing that the mean square linking number of the Brownian loop with a single hoop is infinite. Interestingly, the real part (of (32), coming from the $_2F_2$ in (34)) does not diverge for coincidence, but as remarked earlier, an interpretation of this real part is lacking.

The linking in question has been between the random path and the two hoops, but one might ask whether it matters if the two hoops themselves are linked with each other (or even knotted). It does not matter; there is no discontinuity (of value) of the integral (32) if the two hoop curves are deformed to pass through each other, and the formula continues to apply.

*Appendix: Thermal quantum correlation function*

The two next formulas (A1) and (A2), are (differential) operator renditions, respectively, of a path integral for Brownian motion (), and a path integral for thermal quantum mechanics (). They are evidently the same if $t_{12}= i\tau$. In the present context the two symbols $O$ mean $O_2(\tau) = \delta(\mathbf{r}(\tau) - \mathbf{R})$ and $O_1(0) = \delta(\mathbf{r}(0))$, and the Hamiltonian $H$ is kinetic energy $\propto \nabla^2$. The trace $Tr$ can be interpreted, in the respective path pictures, as the integration over all space of the coincident starting and finishing points of the path; symbolically $\int \langle \mathbf{r} | \quad | \mathbf{r} \rangle d^3\mathbf{r}$. The denominator normalization is the partition function (the prefactor in (7) as remarked). The task here is to justify that each indeed represents its process as claimed, and hence that the two processes are related by the rotation of the delay time through a right angle in the complex time plane.

The first, for Brownian motion, is more or less direct: two successive intervals $t_{12}$ and $t-t_{12}$ of diffusion punctuated by the delta functions constraining position. With $0<t_{12}<t$

$$\frac{Tr\left(Exp[-H(t-t_{12})/\hbar]\, O_2(t_{12})\, Exp[-Ht_{12}/\hbar]\, O_1(0)\right)}{Tr\left(Exp[-Ht/\hbar]\right)} \tag{A1}$$

For thermal quantum mechanics (with $\tau>0$)

$$\frac{Tr\left(Exp[-H(t-i\tau)/\hbar]\, O_2(\tau)\, Exp[-iH\tau/\hbar]\, O_1(0)\right)}{Tr\left(Exp[-Ht/\hbar]\right)} \tag{A2}$$

Following Wen [2004] (though restricting to $\tau>0$), one can work in the spectral (energy eigenstate) representation. Writing this as though the spectrum were discrete ( $H|n\rangle = \varepsilon_n |n\rangle$ ) for simplicity:

$$\frac{\langle n|O_2(\tau)\exp[-iH\tau/\hbar]O_1(0)|n\rangle}{\langle n|\exp[-iH\tau/\hbar]|n\rangle}$$

$$= \frac{\sum_m \langle n|O_2(\tau)|m\rangle\langle m|O_1(0)|n\rangle \exp[-i\varepsilon_m\tau/\hbar]}{\langle n|\exp[-i\varepsilon_n\tau/\hbar]|n\rangle} \quad (A3)$$

$$= \sum_m \langle n|O_2(\tau)|m\rangle\langle m|O_1(0)|n\rangle \exp[-i(\varepsilon_m - \varepsilon_n)\tau/\hbar]$$

So with the partition function being a weighted sum over states *n*:
$Z = \sum_n \langle n|\exp[-\varepsilon_n t/\hbar]|n\rangle$, one has

$$\sum_n \frac{\exp[-\varepsilon_n t/\hbar]}{Z} \sum_m \langle n|O_2(\tau)|m\rangle\langle m|O_1(0)|n\rangle \exp[-i(\varepsilon_m - \varepsilon_n)\tau/\hbar]$$
$$= \frac{1}{Z} \sum_n \sum_m \langle n|O_2(\tau)|m\rangle\langle m|O_1(0)|n\rangle \exp[-i\varepsilon_m\tau/\hbar]\exp[-\varepsilon_n(t-i\tau)] \quad (A4)$$

which is (A2) as claimed.

Acknowledgement
I am grateful to Adam Nahum for useful comments on true versus fake time.

*References*

Derrick, G.H., 1985. A three-dimensional analogue of the Hottel string construction for radiation transfer. *Optica Acta: International Journal of Optics*, *32*(1), pp.39-60.

Feynman RP, Leighton RB, and Sands M, 1964, *The Feynman lectures on physics*, vol 2, eqn 17.30. Addison-Wesley, Reading, Massachusetts.

Feynman RP and Hibbs AR 1965, *Quantum mechanics and Path integrals*, McGraw-Hill, New York.

Feynman RP 1972, *Statistical mechanics*, W.A. Benjamin, Massachusetts.

Hannay, J.H., 1995. Path-linking interpretation of Kirchhoff diffraction. *Proc. R. Soc. Lond. A*, *450*(1938), pp.51-65.

Hannay, J.H., 2001. Path–linking interpretation of Kirchhoff diffraction: a summary. *Philosophical Transactions of the Royal Society of London A: Mathematical, Physical and Engineering Sciences*, *359*(1784), pp.1473-1478.

Hannay, J.H., 2018. (with corrigendum) Winding number correlation for a Brownian loop in a plane. *Journal of Physics A: Mathematical and Theoretical*.

Hannay, J.H., 2019. The correlated linking numbers of a Brownian loop with two arbitrary curves. *Journal of Physics A: Mathematical and Theoretical*, *52*(49), p.495201.

Hottel, H.C., 1954. Radiant heat transmission. *WH McAdams. Heat Transmission*.

Sparrow, E.M., 1963. A new and simpler formulation for radiative angle factors. *Journal of Heat Transfer*, 85, pp.81-88.

Wen, X-G., 2004, *Quantum field theory of many body systems*, Oxford University Press, Oxford.

Wiener, N., 1921. The average of an analytic functional. *Proceedings of the National Academy of Sciences of the United States of America*, *7*(9), p.253.